\documentclass[aps,twocolumn, notitlepage, superscriptaddress, floatfix, prl]{revtex4-1}

\usepackage[utf8]{inputenc}
\usepackage{graphicx,pdfsync,amsmath,appendix, braket}
\usepackage{xcolor, hyperref}

\newcommand{\beq}{\begin{equation}}
\newcommand{\eeq}{\end{equation}}

\newcommand{\vect}[1]{\mathbf{#1}}
\newcommand{\aff}{JILA, National Institute of Standards and Technology and University of Colorado, and Department of Physics, University of Colorado, Boulder, CO 80309, USA}
\newcommand{\affT}{Center for Theory of Quantum Matter, University of Colorado, Boulder, CO 80309, USA}

\begin{document}

\title{Thermodynamics of a deeply degenerate SU($N$)-symmetric Fermi gas}

\author{L. Sonderhouse}
\email[email: ]{lindsay.sonderhouse@colorado.edu}
\affiliation{\aff}
\author{C. Sanner}
\author{R. B. Hutson}
\affiliation{\aff}
\author{A. Goban}
\email[Present address: ]{Waseda Research Institute for Science and Engineering, Waseda University, Tokyo 169-8555, Japan}
\affiliation{\aff}
\author{T. Bilitewski}
\affiliation{\aff}
\affiliation{\affT}
\author{L. Yan}
\author{W. R. Milner}
\affiliation{\aff}
\author{A. M. Rey}
\affiliation{\aff}
\affiliation{\affT}
\author{J. Ye} 
\affiliation{\aff}
\date{\today}

\begin{abstract}
Many-body quantum systems can exhibit a striking degree of symmetry unparalleled by their classical counterparts. While in real materials SU($N$) symmetry is an idealization, this symmetry is pristinely realized in fully controllable ultracold alkaline-earth atomic gases. Here, we study an SU($N$)-symmetric Fermi liquid of $^{87}$Sr atoms, where $N$ can be tuned to be as large as 10. In the deeply degenerate regime, we show through precise measurements of density fluctuations and expansion dynamics that the large $N$ of spin states under SU($N$) symmetry leads to pronounced interaction effects in a system with a nominally negligible interaction parameter. Accounting for these effects we demonstrate thermometry accurate to one-hundredth of the Fermi energy. We also demonstrate record speed for preparing degenerate Fermi seas, reaching $T/T_F = 0.12$ in under 3\,s, enabled by the SU($N$) symmetric interactions. This, along with the introduction of a new spin polarizing method, enables operation of a 3D optical lattice clock in the band insulating-regime. 
\end{abstract}

\maketitle

Modern quantum science strives to understand and harness the rich physics underlying interacting many-particle systems. In particular, ultracold quantum gases have been established as ideal model systems to explore open questions in condensed matter physics ranging from understanding superconductivity to realising highly correlated states and exotic phases of matter \cite{bloch_many-body_2008, ketterle_making_2008}. 
It is often the interplay of interactions and symmetries that shapes the complex behavior in these quantum many-body systems \cite{cazalilla_ultracold_2014}.

The dominant interactions at ultralow temperatures in quantum gases occur via $s$-wave contact scattering, which, for identical fermions, vanishes due to the anti-symmetry of fermionic wavefunctions. 
Therefore, multi-component gases are required to realize interacting Fermi systems. Whereas binary mixtures are well studied versatile platforms for quantum simulation capable of exploring a wide parameter space \cite{horikoshi_measurement_2010, nascimbene_exploring_2010}, the additional freedom in choosing the number of spin components, $N$, offers unique, largely unexplored opportunities.

Indeed, in alkaline-earth fermionic atoms, the complete decoupling between the nuclear and electronic degrees of freedom intrinsic to their internal ground state manifold  gives rise to perfectly SU($N)$-symmetric two-body  interactions, characterized by a single nuclear-spin-independent scattering length $a$.
This raises the question to what extent this enlarged symmetry enriches the many-body behavior, and how it affects the thermodynamic and statistical properties of the quantum gas as $N$ is increased.

\begin{figure}[t!] 
\centerline{\includegraphics[scale=.5]{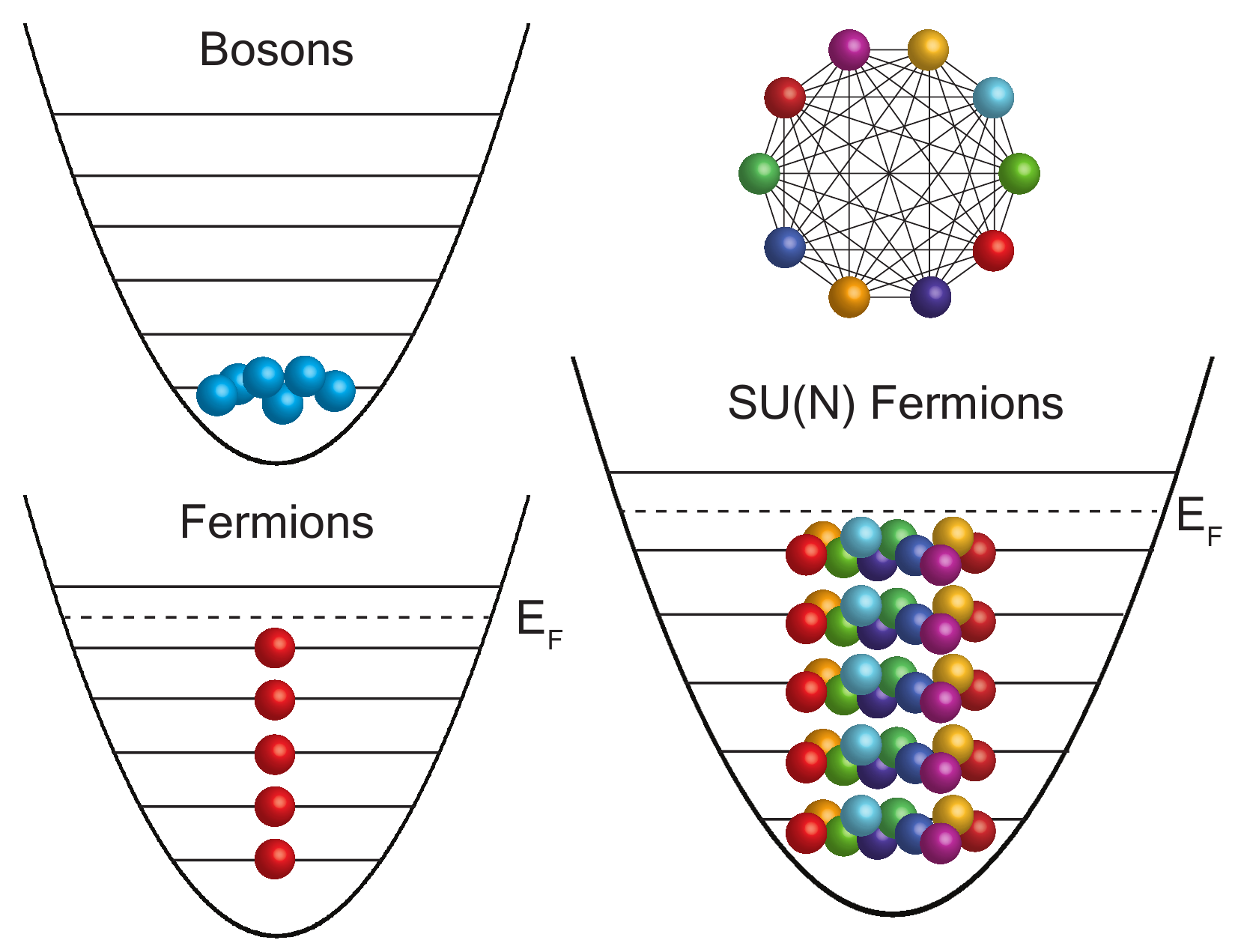}}
\caption{{\bf Degenerate bosons, fermions, and SU($N$) fermions.} Unlike bosons which can occupy the same state, indistinguishable fermions must separate into different energy levels. SU($N$) fermions, on the other hand, can have $N$ particles per state. In a given level, each particle has $(N-1)$ distinct partners, as shown in the top right, and interactions are correspondingly enhanced. }
\label{fig:Schematic}
\end{figure}

While recent studies have started to investigate the intriguing properties of SU($N$) quantum matter, most of the effort so far has been concentrated on the investigation of lattice-confined gases \cite{gorshkov_two-orbital_2010, cazalilla_ultracold_2014, taie_an_2012, scazza_observation_2014, hofrichter_direct_2016, goban_emergence_2018, ozawa_antiferromagnetic_2018}. On the other hand, experiments probing the role of SU($N$) interactions in a regime where a Fermi liquid description is accurate have been limited to non-degenerate \cite{zhang_spectroscopic_2014} or only slightly degenerate gases \cite{pagano_one-dimensional_2014, song_evidence_2020, he_collective_2020,Stellmerbook}. Here, we explore a deeply degenerate regime where $N$ Fermi seas coexist and fundamentally modify the system's thermodynamics (Fig.~\ref{fig:Schematic}).

In addition to enriching the system's quantum behavior, SU($N$) symmetry is an untapped resource for cooling \cite{taie_an_2012,Hazzard2012,Bonnes2012,Messio2012}. It can enhance the collision and thermalization rate during evaporative cooling without inelastic spin collisions and, thus, provide a tool to efficiently remove entropy from the system.

In this work, we study a deeply degenerate SU($N$) symmetric $^{87}$Sr Fermi gas with $N \leq 10$. We show how the large number of nuclear spin states enables an unprecedented short preparation time to reach Fermi degeneracy with a temperature $T/T_F = 0.12$ achieved in just under 3 s, where $T_F$ is the Fermi temperature. We use the rapid preparation and deep degeneracy to study the pronounced modifications introduced by SU($N$) interactions on the thermodynamic properties of the gas \cite{yip_theory_2014}. In particular, we experimentally demonstrate and theoretically validate within a kinetic approach the effects of SU($N$) interactions on the density profile, number fluctuations and compressibility, as well as the time-of-flight dynamics of the gas. Our study paves the way for future investigation of SU($N$) interactions in regimes where even richer physics emerges and Fermi liquid theory becomes invalid. Furthermore, we introduce a Stark-shift-enabled spin selection technique to spin-polarize deeply degenerate gases, allowing efficient preparation of a low entropy Fermi gas that is invaluable for further development of a quantum-degenerate 3D lattice clock \cite{campbell_fermi-degenerate_2017, marti_imaging_2018}. Our spin selection technique improves the stability of state-of-the-art optical clocks, which are currently limited by the time spent preparing the system \cite{oelker_demonstration_2019, mcgrew_atomic_2018, huntemann_single-ion_2016, sanner_optical_2019, kolkowitz_gravitational_2016, xu_probing_2019, schioppo_ultrastable_2017}. This is particularly relevant in the context of recently proposed lattice clocks relying on engineered quantum states of matter \cite{hutson_engineering_2019}.

\begin{figure*}[t!]
\centerline{\includegraphics[width = 5.7 in]{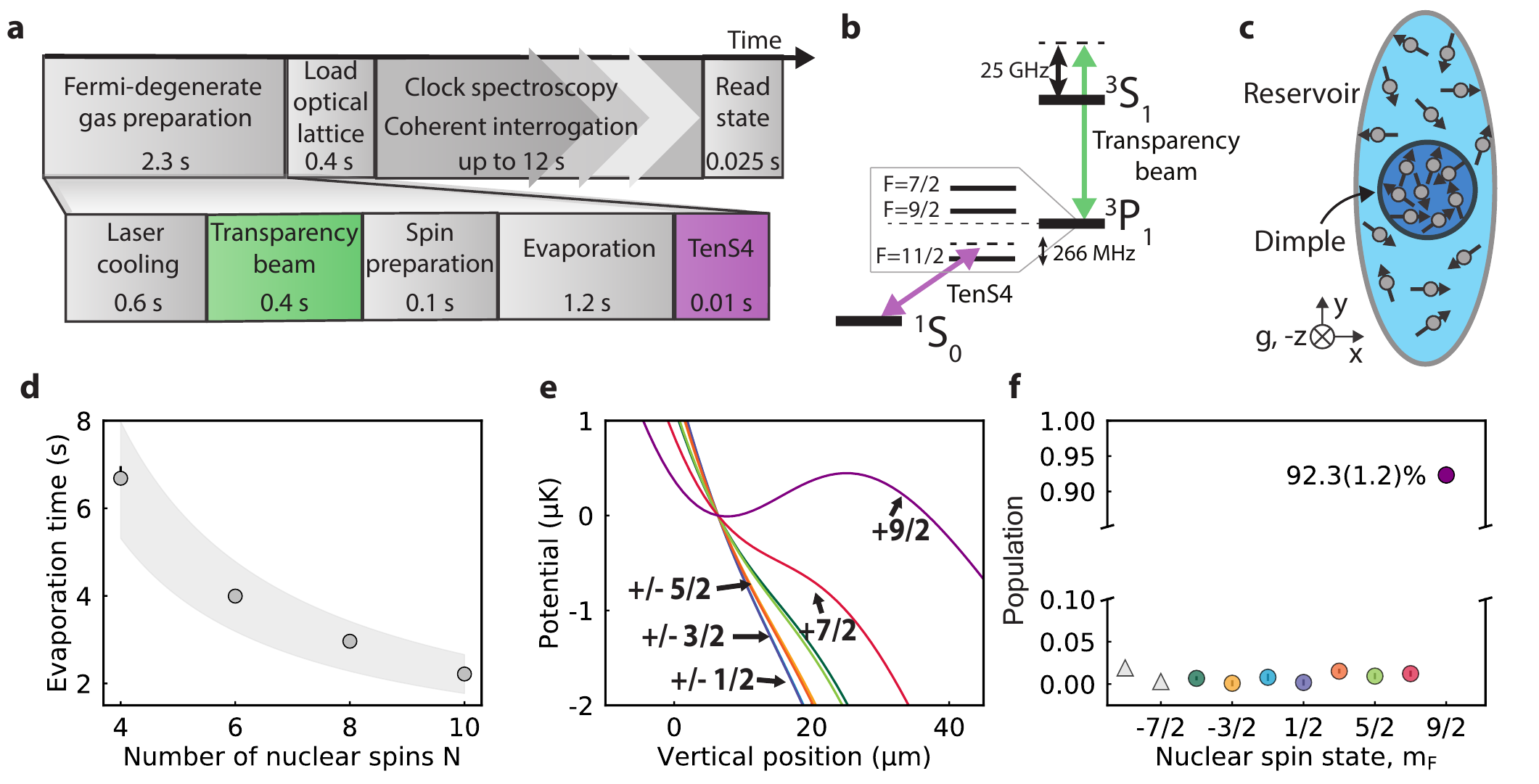}}
\caption{{\bf Preparation of a degenerate Fermi gas.} {\bf a,} Timing diagram showing stages of the experimental sequence. Single spin samples with 20,000 atoms at $T/T_F = 0.2$ are prepared in a 3D optical lattice in under 3\,s. {\bf b,} Level diagram depicting the energy transitions for the transparency beam and Tensor Stark Shift Spin Selector (TenS4). {\bf c,} Atoms are confined in a crossed optical dipole trap consisting of a reservoir and dimple trap, and accumulate in a dimple after a ``transparency'' laser is applied. The transparency and TenS4 lasers overlap with the dimple, and form a $10^{\circ}$ angle with respect to gravity. {\bf d,} Time to reach $T/T_F = 0.12$ for different numbers of nuclear spin states $N$ participating in evaporation. Each trajectory is prepared with $9.6(8) \times 10^5$ atoms per spin state and $T/T_F = 2.0(1)$. The gray band denotes a $1/(N-1)$ scaling based on the number of initial collisional partners, given a $\pm 20\%$ change in the atom number. {\bf e,} Combined optical and gravitational potential after application of the TenS4 laser. The beam creates a spin-dependent modification of the potential, and only atoms in spin state $\mathrm{m_F = + 9/2}$ (purple line) are supported against gravity. Atoms in $\mathrm{m_F} = -9/2$ and $-7/2$ are removed through optical pumping. {\bf f,} Spin purity after application of the TenS4 laser for 10\,ms. Atoms are loaded into a 3D optical lattice and the spin state population is determined using selective excitation on the ultranarrow $\mathrm{^1S_0 – ^3P_0}$ transition. Gray triangles correspond to spin states that are removed through optical pumping. Error bars are included. }
\label{fig:SpinPurity}
\end{figure*}

{\bf Fast preparation} \\
Our preparation scheme (Fig.~\ref{fig:SpinPurity}a) begins with standard laser cooling techniques developed for alkaline-earth atoms \cite{loftus_narrow_2004}. After two stages of laser cooling, roughly $10^7$ atoms are cooled to 2 $\mu$K in a far-off-resonant crossed optical dipole trap (XODT) with a sheet-like geometry \cite{campbell_fermi-degenerate_2017, stellmer_production_2013}. A vertically oriented round optical dipole trap (VODT) forms a dimple in a horizontal optical sheet potential that is provided by an elliptically shaped horizontal optical dipole trap (HODT) (Fig.~\ref{fig:SpinPurity}c). The HODT provides support against gravity and therefore determines the effective trap depth.

The density obtainable in a MOT is generally limited by inelastic light-assisted collisions \cite{julienne_theory_1992} and reabsorption of the cooling light, which leads to an effective repulsion between atoms \cite{sesko_behavior_1991}. To further increase the density inside the dimple beyond these limits, we locally apply an additional laser that renders atoms inside the dimple region transparent to MOT light, a method that was adapted from bosonic $^{84}$Sr \cite{stellmer_laser_2013}. The beam spatially overlaps with the VODT but has a slightly larger waist.  This ``transparency'' laser shifts the cooling light out of resonance, and a high density can then be reached inside the dimple, where atoms collect. Atoms in the dimple thermalize with atoms in a large reservoir part of the trap, which are continually being cooled by the MOT light.

The transparency beam  has allowed the production of  Bose-Einstein condensation without additional evaporation \cite{stellmer_laser_2013}. However, for fermions additional complications can arise. Compared to bosonic strontium, $\mathrm{^{87}Sr}$ has additional hyperfine structure, creating a different AC Stark shift for each nuclear sublevel. Transparency from both the trapping and stirring second stage MOT lasers is also required \cite{mukaiyama_recoil-limited_2003}. Accordingly, we use transparency light that is blue-detuned by 25\,GHz from the $\mathrm{^3P_1 - {^3S_1}}$ transition (Fig.~\ref{fig:SpinPurity}b). This provides an ample shift for all nuclear spin states and keeps spontaneous scattering events beyond relevant experimental time scales. The beam is linearly polarized and  perpendicular to the magnetic quantization field, which is applied along gravity. With this geometry, the smallest Stark shift still detunes the excited state by 250 linewidths. After 400\,ms of cooling with the transparency beam, the number of atoms in the dimple saturates with $5 \times 10^6$ atoms, i.e. about 50\% of the total atom number, a temperature of $2\,\mathrm{\mu K}$, and $\mathrm{T/T_F} = 1.9$ (see Methods). Lower temperatures can be achieved by adjusting the MOT light, but without further improving the phase space density.

The very dense and almost degenerate sample can then be further cooled via forced evaporation. Spin relaxation is absent due to the SU($N$)-symmetric nature of $\mathrm{^{87}Sr}$ in the $\mathrm{^1S_0}$ ground state; however, over the timescale of seconds we observe a decay of the sample which is well described by a 3-body loss process with $k_3 = 4.7(1.2) \times 10^{-30}\, \mathrm{cm^6/s}$ (see Methods).

Before quantum degeneracy, the elastic collision rate \cite{fukuhara_degenerate_2007} for a balanced spin mixture with spatially averaged single-spin density $\bar{n}_\sigma$ is proportional to $(1-1/N)  \,N \bar{n}_\sigma$. Assuming a constant atom number per spin state, it is thus advantageous to have all 10 spin states populated. We reach an initial collision rate of $\mathrm{1000\,s^{-1}}$ with $N = 10$. Evaporation begins at a trap depth of $20\,\mathrm{\mu K}$ with trap frequencies in the dimple of $(\mathrm{\nu_r, \nu_z}) = (100, 800) \, \mathrm{Hz}$. The HODT intensity is then reduced in a two-stage ramp down to a final trap depth of a few 100\,nK with trap frequencies of $(100, 200) \,\mathrm{Hz}$.

After 600\,ms of evaporation, we reach $T/T_F = 0.22$ with $3 \times 10^4$ atoms per spin state. Slower evaporation leads to lower temperatures, and we achieve $T/T_F = 0.07$ with $5 \times 10^4$ atoms per spin state after evaporating for 2.4\,s. This marks a considerable improvement over previous evaporation results, where evaporation stages took around 10\,s \cite{campbell_fermi-degenerate_2017}, limiting the full potential of Fermi-degenerate optical atomic clocks \cite{dick_local_1987}.

We observe an approximate $1/(N-1)$ scaling of the total evaporation time with the number of spin states participating in evaporation as shown in  Fig.~\ref{fig:SpinPurity}d, reflecting the reduction in collisional partners for smaller $N$. Here, each sample is prepared with the same atom number per spin state and $T/T_F$, and is measured after reaching $T/T_F = 0.12$. The final atom number per spin state changes by roughly a factor of two as $N$ varies.

{\bf Spin Manipulation} \\
In order to manipulate the spin composition of the atom sample, particularly to prepare a spin-polarized gas, we apply a spin-selective optical potential to the atoms after evaporation. While past procedures have used optical Stern-Gerlach techniques to separate out spin states during time-of-flight \cite{sleator_experimental_1992, taie_realization_2010, stellmer_detection_2011}, our method, the Tensor Stark Shift Spin Selector (TenS4), creates a spin-selective force on the atoms from the tensor Stark shift of a laser while the atoms remain trapped in the XODT. Atoms with the same $|\mathrm{m_F}|$ experience a small differential force due to an applied magnetic field of 5\,G. This magnetic field is too small to fully remove $\mathrm{m_F} = -9/2$ and $-7/2$. We thus conventionally remove these spins via optical pumping prior to evaporative cooling. For the SU($N$) measurements in the following section, however, we always use all 10 nuclear spins. The TenS4 beam is offset from the atoms such that the AC Stark shift varies across the atomic sample by hundreds of nK, which causes a spin-dependent modification in the combined optical and gravitational potential of the atoms (Fig.~\ref{fig:SpinPurity}e). The TenS4 laser is blue-detuned from the $\mathrm{^3P_1}$, $\mathrm{F=11/2}$ transition by 266\,MHz (Fig.~\ref{fig:SpinPurity}b), where the polarizability from $^3P_1$, F = 11/2 cancels the polarizability from $\mathrm{^3P_1}$, F = 9/2 for nuclear spin state $\mathrm{m_F} = +9/2$. As a result, atoms with $\mathrm{m_F} = +9/2$ are unaffected by the TenS4 laser while all other spin states feel a repelling force. A detailed experimental protocol is provided in the methods.

The spin purity after applying the TenS4 laser for 10\,ms is measured by loading the atoms into a deep 3D optical lattice. The spin population for each nuclear spin state is then read out through selective $\pi$-pulse excitations on the clock transition. We measure 92\% of the atoms in the target $\mathrm{m_F} = +9/2$ state, as shown in Fig.~\ref{fig:SpinPurity}f, and an atom number of $3.3 \times 10^4$ after application of the TenS4 beam, in rough agreement with $\mathrm{1/8^{th}}$ of the initial atom population of $2.5 \times 10^5$. The temperature of the sample heats by only $\sim$10\%, producing $\mathrm{T/T_F} = 0.2$. Our technique provides spin-state selectivity without optical excitation and as a result does not cause light-induced heating, overcoming issues typically associated with optical pumping schemes.
\begin{figure}[t!]
\centerline{\includegraphics[scale=.55]{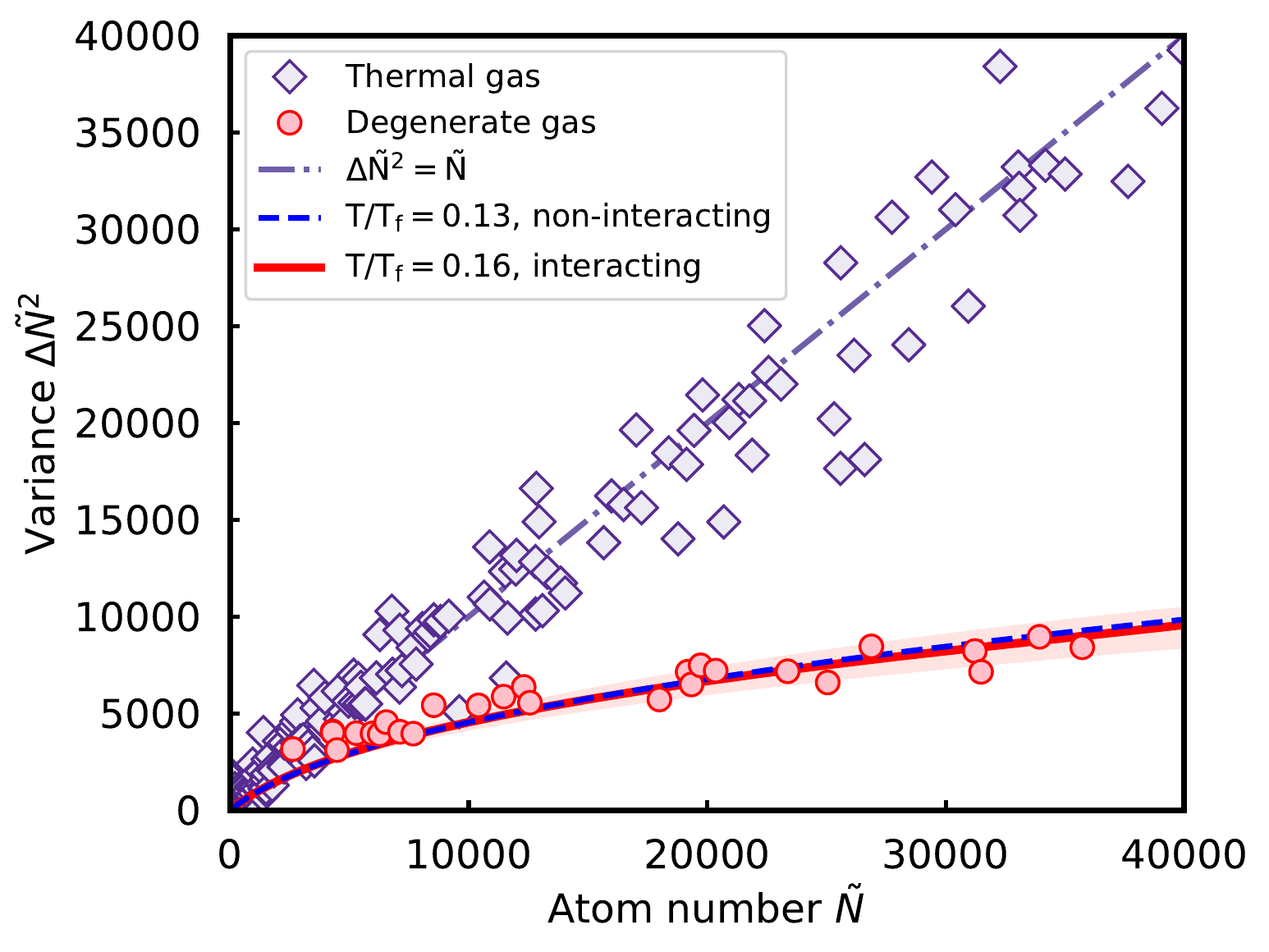}}
\caption{{\bf Local density fluctuations of an SU($N$) degenerate gas.} Density fluctuations after 11.5\,ms time-of-flight for a degenerate cloud with $N=10$ nuclear spin states (red points). The data is fit using an SU($N$) interacting model to extract $T/T_F=0.16$ (red solid line), with shading representing a $2 \sigma$ uncertainty of $\pm 0.02 \,T/T_F$. Fitting the data instead to a non-interacting ideal Fermi gas gives $T/T_F=0.13$ (blue dashed line), showing an interaction-induced suppression of $\sim$20\%. The total density fluctuations are 25\% of that of the thermal gas. A thermal cloud (purple points) reproduces Poisson statistics with $\Delta \tilde{N}^2/\tilde{N} = 1$ (purple dot-dashed line).}
\label{fig:3}
\end{figure}

{\bf Characterization of SU($N$)-enhanced interactions} \\ Having prepared a high-density deeply degenerate SU(10) gas, we demonstrate in the following that a nominally very weakly interacting quantum system with interaction parameter $k_F a \ll 1$, where $k_F$ is the Fermi wave vector, can develop striking interaction effects due to SU($N$) enhancement.  The nuclear spin degree of freedom substantially modifies the character of the gas towards an interacting multi-component Fermi liquid with subtle consequences for correlation analysis and thermometry.
\begin{figure*}[t!]
\centerline{\includegraphics[scale=.7]{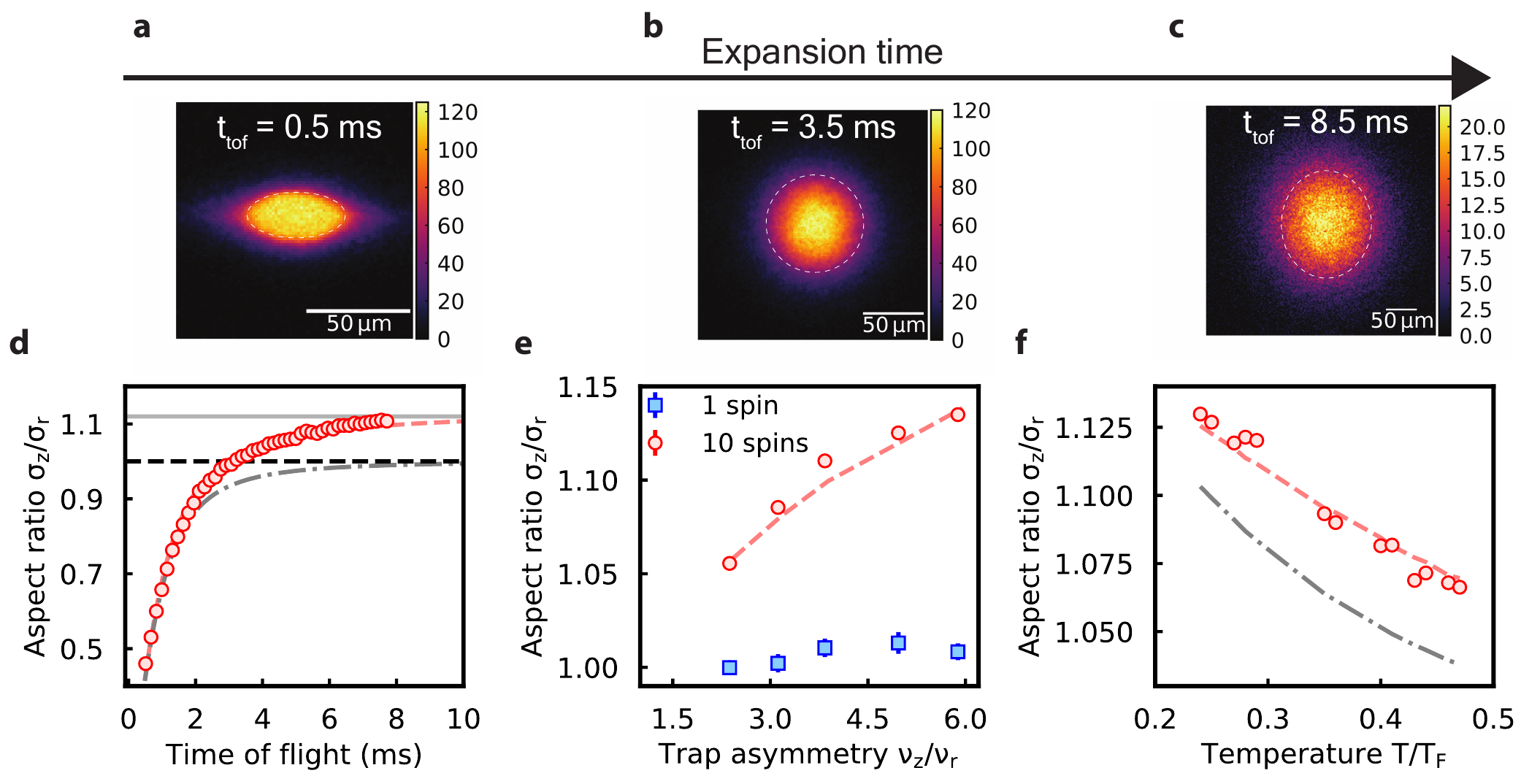}}
\caption{{\bf Cloud anisotropy.} {\bf a, b, c,} Line-of-sight integrated atomic density for 0.5 ms, 3.5 ms, and 8.5 ms time-of-flight ($t_{tof}$) expansion times. The colorbar's unit corresponds to the number of atoms per 1.37 $\mathrm{\mu m}^2$. {\bf d,} Aspect ratio of a cloud of cold atoms with 10 nuclear spin states released from an optical dipole trap for variable expansion times. After $\sim$3\,ms, the aspect ratio passes through unity (black dashed line), a clear signature of interactions in the gas. At long times, the sample approaches an aspect ratio of 1.12 (grey line). The sample has an initial trap asymmetry of $\mathrm{\nu_z/\nu_r = 6.4}$. The red dashed line shows the expected aspect ratio with $T/T_F = 0.16$ assuming an interacting model; without interactions the theory produces the grey dash-dotted line. {\bf e,} Aspect ratio  versus initial trap asymmetry of a degenerate gas for 10 (red circles) and 1 (blue squares) spin states after time-of-flight expansion for 15.5\,ms. The interacting model reproduces the experimental results (red dashed line). {\bf f,} Aspect ratio versus temperature. Data is shown with roughly the same atom number per shot. The data is fit using an interacting model that includes both a mean-field interaction and an additional collisional term (red dashed). Neglecting the collisional term fails to explain the results (gray dash-dotted). }
\label{fig:4}
\end{figure*}

In order to investigate this intriguing quantum system and illuminate the role of SU($N$) symmetry, we perform measurements that characterize the system's thermodynamics. A key quantity in this context is the isothermal compressibility $\kappa = \frac{1}{n^2} \frac{\partial n}{\partial \mu}$, where $n = N  \, n_\sigma$ denotes the particle density and $\mu$ the chemical potential. For $^{87}$Sr with $a=97 a_{\mathrm{Bohr}}$, the contact interactions are repulsive and one expects a decreased compressibility compared to an ideal Fermi gas with compressibility $\kappa_0$.  For a homogeneous gas in the zero temperature limit where $\kappa_0 = 3/(2 N n_\sigma E_F)$ with Fermi energy $E_F$ and Fermi wave vector $k_F = (6 \pi^2 n_\sigma)^{1/3}$,  one finds to first order in $k_F a$ the ratio $\kappa_0 / \kappa = 1 + (N-1) \, 2 k_F a / \pi$ \cite{yip_theory_2014}. Therefore, the symmetric $N$-component system is effectively $(N-1)$-fold more repulsive than a typical two spin component Fermi liquid. The favorable scaling with number of internal levels has to be contrasted with the weak dependence of the compressibility on the atom number per spin state in a harmonically trapped gas, where $k_F \propto N_\sigma^{1/6}$.

Experimentally, we access the compressibility of the gas by measuring its local density fluctuations. Fluctuations, either thermal or quantum, are the drivers of phase transitions, and are sensitive to the underlying phase of matter, its quasi-particles and interactions. The fluctuation-dissipation theorem states that the thermally driven fluctuations of a thermodynamic variable are fundamentally related to the conjugate external force through the susceptibility \cite{callen_irreversibility_1951}. Considering a small subvolume of the gas cloud containing on average $\tilde{N}$ atoms, the corresponding generalized force is the local chemical potential $\mu$. The relative number fluctuations $\eta = \Delta \tilde{N}^2 / \tilde{N}$ are therefore related to the susceptibility $\partial \tilde{N} / \partial \mu$ via $\eta = n k_B T \kappa$, where $k_B$ is the Boltzmann constant. While the equation of state of a classical ideal gas dictates $\eta = 1$, one expects for a deeply degenerate ideal Fermi gas $\eta = 3/2 \; T/T_F$. These sub-Poissonian fluctuations reflect the degeneracy pressure in the gas. Combined with the compressibility reduction due to repulsive interactions one therefore expects to first order in temperature and interactions (see Methods):
\begin{equation*}
\eta = \frac{3}{2} \frac{T/T_F}{1+\frac{2}{\pi} \,(k_F a) (N-1)}\,,
\end{equation*}
which suggests that even in the $k_F a \ll 1$ limit the interaction effects become non-negligible due to the $(N-1)$-fold SU($N$) enhancement.

Our density fluctuation measurements are performed on expanded gas clouds.
After abruptly turning off the harmonic confinement ($\nu_r=130 \, \mathrm{Hz}, \nu_z= 240 \, \mathrm{Hz}$) the quantum degenerate sample that contains in total $10 \times 59000$ atoms such that $k_F a = 0.07$ freely expands over 11.5 ms. We then obtain line-of-sight integrated density profiles via absorption imaging. Following the protocol described in \cite{sanner_suppression_2010} we run this experiment in a repeated fashion so that for each projected subregion of the cloud containing on average $\tilde{N}$ atoms we can measure the statistical variance $\Delta \tilde{N}^2$. Fig.~\ref{fig:3} shows the results obtained from 400 individual images together with a calibration line derived from noise measurements on a thermal gas. Pronounced noise suppression down to about 25\% of thermal noise in the center of the sample indicates that the gas is deeply in the quantum regime. 

To quantitatively interpret the noise data beyond first order and decouple Pauli suppression and SU($N$)-enhanced interaction contributions, we calculate the expected line-of-sight integrated number fluctuations based on a kinetic approach \cite{Quantum_Kinetic_Theory,kadanoff_quantum}, using the collisional Boltzmann-Vlasov equation with a mean field interaction term. The Boltzman-Vlasov equation describes the evolution of the semi-classical phase-space distribution $f(\vect{r},\vect{p})$:
 \begin{equation}
  \left( \partial_t + \frac{\vect{p}}{m} \cdot \nabla_{\vect{r}}  -\nabla_{\vect{r}} \left[U(\vect{r})+ V_{\mathrm{MF}}(\vect{r}) \right] \cdot \nabla_{\vect{p}} \right)  f = I_c(f) \, ,
  \label{eq:1}
  \end{equation}
due to ballistic motion of particles (second term), the forces due to the harmonic trapping potential, $U= m/2 \sum_i (2 \pi \nu_i)^2 r_i^2$, the mean field interactions, $V_{\mathrm{MF}} = g (N-1) n$ with $g=4\pi \hbar^2 a/m$, and the collisional integral, $I_c(f)$ (see Methods) \cite{kadanoff_quantum, Stringari_relaxation_time}. Solving the Boltzmann-Vlasov equation in equilibrium and for finite temperature allows us to obtain the real space density, $n$, from which we can compute the compressibility, and thus the number fluctuations in trap and after time of flight. By fitting this model to the observed fluctuations, we extract a temperature $T/T_F = 0.16 \pm 0.01$. At these low temperatures, $I_c(f)$ plays no role.

To illustrate the interaction-induced compressibility change we additionally fit a non-interacting model to the noise data, which gives an apparent $T/T_F = 0.13 \pm 0.01$, indicating a $\sim$20\% compressibility reduction due to interactions. More precisely, we find that the compressibility in the center of the trap is reduced by 18\% compared to a non-interacting gas at the same density and temperature. This percentage is comparable to the ratio of 21\% between the interaction energy in a small volume $V$ at the centre of the cloud, $g/2 \, (N-1) n_{\sigma}^2 V$, and the total energy of a noninteracting Fermi gas at the same density, $3/5 \, E_F n_{\sigma} V$.
Clearly, without prior knowledge of the interaction parameter, one cannot distinguish between a colder or more repulsively interacting system. Having a full thermodynamic description at hand we also perform global profile fits of the acquired images to numerically calculated density distributions and find $T/T_F = 0.17 \pm 0.01$, which is in good agreement with the temperature derived from density noise measurements.

To unambiguously distinguish between temperature and interaction effects we study the expansion dynamics of the cloud after being released from the trap. In addition to the kinetic energy of the gas, interactions provide an additional release energy that is mapped to momentum after long time-of-flight. Since the mean field energy preferentially pushes atoms along the direction of the largest density gradient, this conversion produces an anisotropic distribution after long time-of-flight, see Fig.~\ref{fig:4}a-c. The expansion can be described via scaling solutions of the time-dependent Boltzmann-Vlasov equation (see Methods) \cite{Review_FermiGases,Fermi_Expansion,BCE_Expansion,MF_trapped_gas}. Whereas the effect on the initial density can be partly captured by a lower temperature, the interaction terms during expansion cannot, and results in a non-unity aspect ratio at long times. This is in stark contrast to the expansion behavior of an ideal Fermi gas where expansion is purely ballistic and after long time-of-flight reflects the isotropic momentum distribution even if the confining potential is anisotropic.

Fig.~\ref{fig:4}d displays the aspect ratio of a SU(10) atom cloud measured after variable expansion times $t_{\mathrm{tof}}$ out of a harmonic trap with $\nu_r=125$ Hz and $\nu_z=800$ Hz. The sample contains 50,000 atoms per spin component at $T/T_F = 0.16$. Initially, the atom cloud reproduces the trap's asymmetry. As $\nu_{r, z} t_{\mathrm{tof}}$ becomes larger than one, the spatial density distribution is more and more determined by the momentum distribution in the gas. Observing an inversion of the aspect ratio beyond 1 is an unambiguous signature that interactions have modified the isotropic momentum distribution during time-of-flight \cite{ohara_observation_2002}.

To further explore this behavior we present in Fig.~\ref{fig:4}e measurements of the expanded cloud aspect ratio ($t_{\mathrm{tof}} = 15.5$ ms) as a function of the confinement asymmetry for a 10-component gas and a spin-polarized gas, both at $T/T_F =0.16$. In the non-interacting case $N = 1$ (blue data points), the aspect ratio is always 1 as expected in the long time-of-flight limit. Finally, we show in Fig.~\ref{fig:4}f the dependence of the observed cloud aspect ratio on $T/T_F$ for a fixed initial confinement with $\nu_r=130$ Hz and $\nu_z=725$ Hz and unchanged $t_{\mathrm{tof}} = 15.5$ ms. Atomic interactions include a mean field term $\propto a$ and a collision integral $I_c(f)\propto a^2$, see equation (\ref{eq:1}). The latter, however, is only relevant in the presence of high collision rates, and, in fact, its pronounced effect is observed when the gas is relatively hot (Fig.~\ref{fig:4}f). In comparison, at low temperatures the interaction energy and kinetic energy become comparable, and the collisional rate is suppressed by Fermi statistics.

All measurements are well reproduced by our quantitative model (see Methods). Because aspect ratio measurements via Gaussian fits are fairly immune to most imaging artifacts, they can be exploited to perform precise thermometry of the interacting Fermi gas.

\begin{figure}[t!]
\centerline{\includegraphics[scale=0.9]{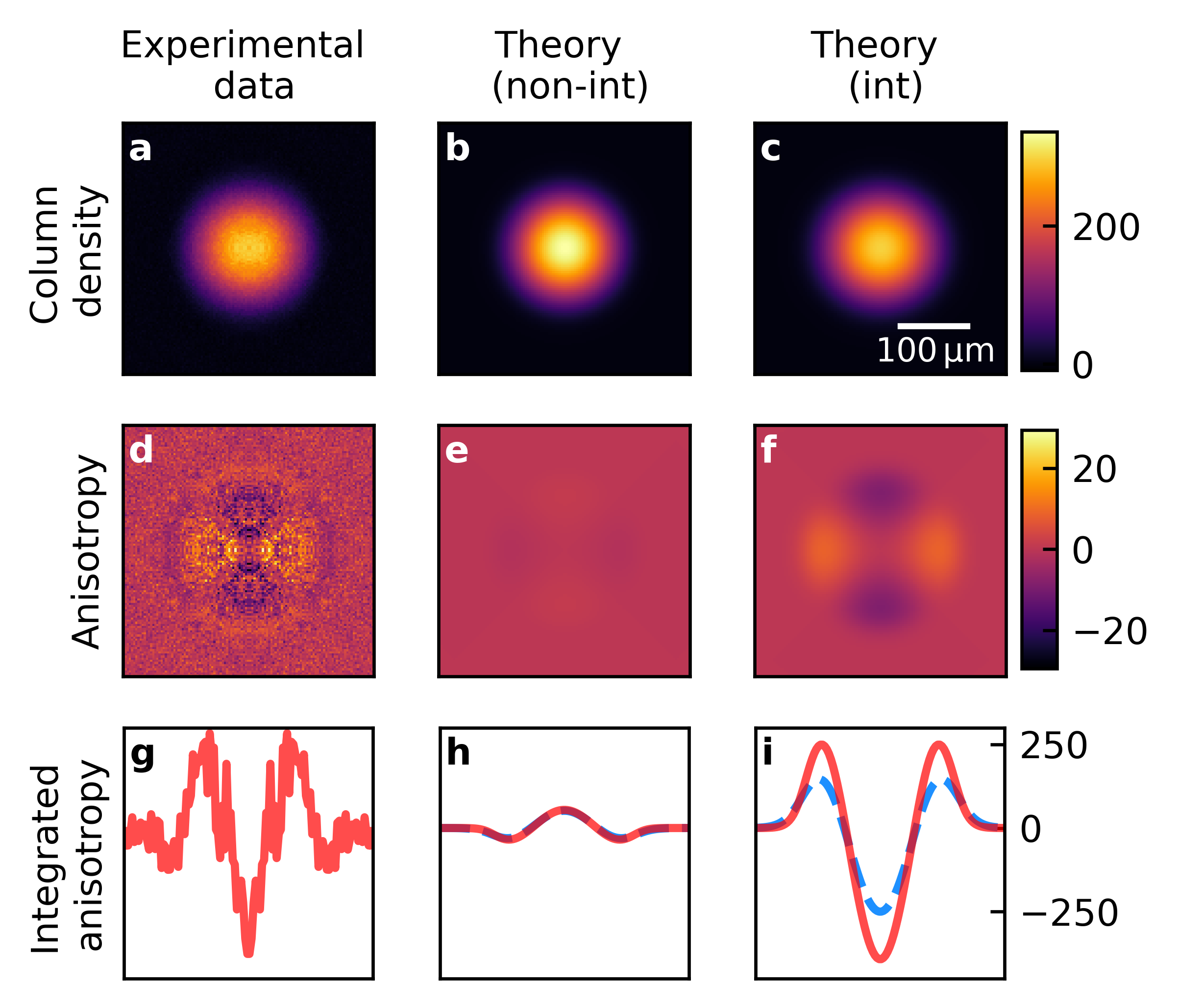}}
\caption{{\bf Interaction signatures.} {\bf a,b,c,} Line-of-sight integrated atomic density $n(x,z)$ after time-of-flight for the experimental data, non-interacting theory, and interacting theory, respectively. The colorbar's units represent the number of atoms per 1.37 $\mathrm{\mu m}^2$. Images are shown with an initial trap asymmetry of $\nu_z/\nu_r = 1.8$ and $T/T_F = 0.17$. {\bf d,e,f,} The anisotropy of the cloud, defined as $n(x,z)-n(z,x)$, is shown in the second row, where lobes are clearly visible for clouds with interactions. To improve the signal-to-noise ratio the experimental image is first symmetrized by reflection along the $x$ and $z$-axes. {\bf g,h,i,} If the anisotropy is integrated along one direction, peaks symmetric to the center of the gas appear for the interacting distribution that are sensitive to temperature, while the non-interacting signal shows a different signature that displays only weak temperature dependence. Here, the red lines show the integrated anisotropy of the images in (d-f), while the blue dashed lines in (h) and (i) show the integrated profile for a 50\% hotter temperature. }
\label{fig:5}
\end{figure}

Beyond their directly visible manifestation through cloud ellipticity one can also identify interaction-modified expansion dynamics by carefully inspecting high-signal-to-noise absorption images. Fig.~\ref{fig:5} illustrates that a presumably round line-of-sight integrated density profile $n(x,z)$ still contains a systematic interaction signature.
After a time-of-flight time of 11.5 ms we observe density profiles that at first sight appear circularly symmetric (first row) for the experimental data (first column), the noninteracting (second column) and the interacting model (third column).
Interactions are revealed in the transpose-anisotropy of the density distribution, defined as $n(x,z)-n(z,x)$, shown in the second row. Both the experimental data and the interacting model exhibit pronounced lobes that are not visible in the non-interacting case. Finally, a one-dimensional measure of this anisotropy can be defined by integrating over one of the axes ($\int dz ( n(x,z)-n(z,x))$) as shown in the third row. In the integrated transpose-anisotropy we observe peaks symmetric to the centre of the cloud whose height are sensitive to $T/T_F$ for the experimental data and the interacting model, whereas in the noninteracting case this anisotropy is significantly reduced, inverted and insensitive to temperature. Thus, the anisotropy reveals the interacting nature of the Fermi gas in a single absorption image and can serve as a precise temperature probe.

{\bf Concluding remarks}\\
We have demonstrated that SU($N$) symmetry substantially enhances interaction dynamics in a quantum degenerate Fermi gas. This allows us to reach ultralow temperatures in record speed. Creating a spin-polarized sample in under 3\,s is an important milestone for the realization of atomic clocks probing engineered quantum states of matter \cite{hutson_engineering_2019}. The many-body problem for the dilute repulsively interacting Fermi gas can be solved exactly, and we have shown with high precision how the additional spin degree of freedom systematically modifies thermodynamic properties in the bulk gas. This opens the path for future quantum simulators to systematically explore SU($N$)-symmetric Fermi systems in periodic potentials.

\bibliographystyle{naturemag}
\bibliography{BibTeX_Library}

\noindent\textbf{Acknowledgments} We are grateful to T. Bothwell, C. Kennedy, E. Oelker, and J. Robinson for discussions and technical contributions. We also thank J. Thompson and C. Kennedy for reading the manuscript. This work is supported by NIST, DARPA, AFOSR grants FA9550-19-1-0275 and FA9550-18-1-0319, and NSF grant PHYS-1734006. C. Sanner thanks the Humboldt Foundation for support.
\\
\noindent\textbf{Author contributions} L.S., C.S., R.B.H., A.G., L.Y., W.M. and J.Y. contributed to the experimental measurements. T.B. and A.M.R. developed the theoretical model. All authors discussed the results, contributed to the data analysis and worked together on the manuscript.
\\
\noindent\textbf{Author information} The authors declare that they have no competing financial interests. Correspondence and requests for materials should be addressed to L. Sonderhouse
(lindsay.sonderhouse@colorado.edu) or J. Ye (Ye@jila.colorado.edu).

\textbf{Methods}
\\
\textbf{Transparency laser:} The transparency light is derived from an extended cavity diode laser (ECDL) that is filtered by a volume Bragg grating. Due to amplified spontaneous emission (ASE), we see a lifetime in the dimple of 5\,s. The beam is linearly polarized along $x$, while a small magnetic field is applied along $z$. To separately extract the number of atoms in the dimple and reservoir, the HODT is extinguished. Atoms in the dimple are guided by the VODT which has a small angle with respect to gravity, while atoms in the reservoir undergo free expansion. This spatially separates the atoms in the dimple and the reservoir, and atom numbers can be respectively counted through absorption imaging. $T/T_F$ is then extracted for atoms in the dimple by calculating $T_F$ and measuring the temperature, which is determined by fitting the reservoir and dimple atoms to a Gaussian fit after extinguishing the XODT after long time-of-flight. \\
\textbf{Three body loss:} To determine the three-body loss coefficient $k_3$, we load a thermal gas with temperature $T = 1.45\,\mathrm{\mu K}$ into the dimple part of the recompressed dipole trap. Starting from an initial central density of $n = 3.9 \times 10^{14}$ cm$^{-3}$ we measure over the next ten seconds a decay of the total atom number $\tilde{N}(t)$ as a function of the holding time $t$. The observed atom loss is modeled as
\begin{equation}
\begin{split}
    \tilde{N}(t)= \tilde{N}(0)&- k_3\int_{0}^{t}d\tau\int{n^3(\textbf{r},\tau)}dV\,,
    \end{split}
\end{equation}
where $n(\textbf{r},t)$ is the atomic density at position \textbf{r} and time $t$. We find a three-body loss coefficient of $k_3 = 4.7(1.2)\times 10^{-30}\,\mathrm{cm^6/s}$. Interestingly, this is a factor of 2 larger than the recent lattice-based 3-body measurement that showed agreement with a universal Van der Waals model \cite{goban_emergence_2018}. Discrepancies in 3-body loss between bulk gas measurements and predictions have been seen before \cite{PhysRevLett.79.337,Wolf921}, suggesting the challenge of accounting for inhomogeneous density profiles. Under our experimental conditions, single and two-body contributions are expected to be negligible over a time interval of ten seconds and using a corresponding model we find them to be statistically insignificant. \\
\textbf{TenS4 laser:} The TenS4 laser overlaps with the VODT and the transparency beam, and has a 30\,$\mathrm{\mu m}$ waist with a peak intensity of $0.15\,\mathrm{kW/cm}^2$. The beam is linearly polarized along $y$, along which a small magnetic field of 5\,G is also applied, producing a small differential force between $\mathrm{m_F}$ states of opposite signs. Due to our relatively small magnetic field and optical power, the TenS4 beam does not provide enough force to completely remove all other nuclear spin states. Consequently, we conventionally remove atoms with $\mathrm{m_F} = -9/2$ and $\mathrm{m_F} = -7/2$ prior to evaporation using optical pumping to aide with spin selectivity. To ensure that we are not addressing a molecular resonance, photoassociation spectra were measured at a variety of detunings from the $\mathrm{^3P_1}$, $\mathrm{F = 11/2}$ resonance. Additionally, a low-finesse cavity is used to filter out ASE from the ECDL laser source which produces significant on-resonant scattering. \\
\textbf{Model:} Our system is well described by the Hamiltonian
\begin{equation}
\begin{split}
  \mathcal{H}  &= \sum_{\sigma}\int d^3\vect{r} \, \hat{\psi}^{\dagger}_{\sigma}(\vect{r}) \left(  -\frac{\nabla^2}{2m}  +U(\vect{r}) \right) \hat{\psi}_{\sigma}(\vect{r})\\
  &+\frac{g}{2}\sum_{\sigma,\sigma^{\prime}}\int d^3\vect{r}  \hat{\psi}^{\dagger}_{\sigma}(\vect{r}) \hat{\psi}_{\sigma^{\prime}}^{\dagger}(\vect{r})\hat{\psi}_{\sigma^{\prime}}(\vect{r}) \hat{\psi}_{\sigma}(\vect{r})
\end{split}
\end{equation}
where $\hat{\psi}_{\sigma}(\vect{r})$ creates a fermion of spin $\sigma$ at point $\vect{r}$. The fermions are confined by a harmonic trapping potential $U= m/2  \sum_i (2 \pi \nu_i)^2
x_i^2$ and interact via SU($N$)-symmetric s-wave contact interactions of
strength $g= 4 \pi \hbar^2 a/m$ given in terms of the scattering length $a$ and the atomic mass $m$.\\
\textbf{Kinetic Approach:} To study the collective behaviour of the quantum gas, we use a kinetic approach \cite{Quantum_Kinetic_Theory,kadanoff_quantum} considering the semi-classical phase space distribution
 $f_{\sigma,\sigma^{\prime}}(\vect{r},\vect{p})$ defined as
 \begin{equation}
 f_{\sigma,\sigma^{\prime}}(\vect{r},\vect{p}) = \int d^3\vect{r}^{\prime} e^{i \vect{p} \vect{r}^{\prime}} \left<\hat{\psi}^{\dagger}_{\sigma} (\vect{r} + \vect{r}^{\prime}) \hat{\psi}_{\sigma^{\prime}}(\vect{r}-\vect{r}^{\prime})\right>
 \end{equation}
in terms of the fermionic creation/annihilation operators. The real space atomic
density is then given by $n_{\sigma}(\vect{r}) = \int d^3p \,
f_{\sigma,\sigma}(\vect{r},\vect{p})$.

We will further assume that the Wigner function is purely diagonal in the spin
indices, i.e. we will neglect any off-diagonal contributions, and since the
experiment prepares an equal number of atoms in each spin state, that the
diagonal terms are all equal, $f_{\sigma,\sigma^{\prime}} = \delta_{\sigma,\sigma^{\prime}} f$.

Finally, we will treat the interparticle interaction in the mean-field
approximation assuming an interaction energy
\begin{equation}
V_{\mathrm{MF}} = g (N-1) n(\vect{r})
\end{equation}
which is directly proportional to the real space density $n(\vect{r})$, the
interaction strength $g$, and the number of spin species $N-1$.

With these approximations the collisional Boltzmann-Vlasov equation reads
 \begin{equation}
  \left( \partial_t + \frac{\vect{p}}{m} \cdot \nabla_{\vect{r}}  -\nabla_{\vect{r}} \left[(N-1)g n(\vect{r})+ U(\vect{r}) \right] \cdot \nabla_{\vect{p}} \right)  f = I_c[f] \, ,
  \end{equation}
which describes the dynamics of the phase-space distribution due to ballistic motion of particles (second term), the forces due to the mean field energy (third term) and the trapping potential (fourth term), and collisions between particles via the collision integral $I_c$ \cite{kadanoff_quantum}. \\
\textbf{Relaxation time approximation:}
Instead of a detailed treatment of the collisions between particles we employ the relaxation time approximation \cite{Stringari_relaxation_time}, which approximates
\begin{equation}
    I_c[f] = -\frac{f -f_{le}}{\tau}\,,
\end{equation}
where $f$ approaches the local equilibrium state $f_{le}$ over a characteristic relaxation time $\tau$.

We follow \cite{Vichi_relaxation_time,Jackson_relaxation_time} and approximate the relaxation time as
\begin{equation}
    (2\pi \nu \tau)^{-1} = (N-1)\frac{4}{5\, 3^{1/3} \pi } \left(\tilde{N}^{1/3} \frac{a}{a_{ho}}\right)^{2} F_Q(T/T_F)\,,
\end{equation}
where $F_Q(T/T_F)$ is a universal function given by an integral \cite{Vichi_relaxation_time}. $F_Q$ vanishes as $(T/T_F)^2$ at low temperatures, a signature of Fermi statistics, is of order 1 in the intermediate temperature regime, and vanishes as $(T/T_F)^{-1}$ at higher temperatures.\\
\textbf{Equilibrium Solution:} Solving the Boltzman-Vlasov equation in equilibrium we obtain 
\begin{equation}
f(\vect{r},\vect{p}) = \frac{1}{e^{\beta \left(\frac{p^2}{2m} + U(\vect{r}) + g (N-1) n(\vect{r}) -\mu \right)}+1}
\end{equation}
which has to be solved self-consistently as it depends on $n(\vect{r})= \int
\frac{d^3\vect{p}}{(2\pi \hbar)^3} f(\vect{r},\vect{p})$. 

In practice, we start by ignoring the interaction term, and determine $\mu_0$
from $\int{\frac{d^3rd^3p}{(2\pi \hbar)^3}} f(\vect{r},\vect{p})=N_{\sigma}$ where $N_{\sigma}$
is the number of atoms per spin species which defines $f_0$. 
Having a $f_{i-1}$ we compute the density $n_i$ and the corresponding interaction
energy $V_{MF,i}$. This defines $f(\mu_i,V_{MF,i})$ from
which we determine $\mu_{i}$ via normalisation.
We then set $f_{i} = \alpha f(\mu_{i},V_{MF,i}) + (1-\alpha) f_{i-1}$ with $\alpha=0.9$. Iterating this procedure leads to convergence in 5-10 iterations for the
parameters we consider.\\
\textbf{Number Fluctuations:} We can compute the expected number fluctuations
directly from the obtained density profiles.
The fluctuation-dissipation theorem describes how the number fluctuations of the gas are related to
the isothermal compressibility:
$\Delta \tilde{N}^2/\tilde{N}=n k_b T \kappa$, where the isothermal compressibility $\kappa = \frac{1}{n^2} \frac{\partial n}{\partial \mu}$.

In the harmonic trap the chemical potential varies as $\mu(\vect{r}) = \mu - U(\vect{r})$. Thus, the derivative with respect to $\mu$ can be
replaced by a derivative with respect to one of the spatial directions as
$\frac{\partial n}{\partial \mu} = \frac{-1}{m (2 \pi \nu_i)^2 r_i} \frac{\partial n}{\partial r_i}$, which we can directly evaluate based on the computed
density profiles. \\
\textbf{Dynamics:} To obtain the dynamics we use the scaling factor method \cite{Review_FermiGases,Stringari_relaxation_time,Fermi_Expansion,BCE_Expansion,MF_trapped_gas} with the ansatz
\begin{equation}
f(\vect{r},\vect{p},t) = \frac{1}{\prod_j (\lambda_j \theta_j^{1/2})}f_0\left(\frac{r_i}{\lambda_i},\frac{1}{\theta_i^{1/2}}(p_i -m\dot{\lambda}/\lambda_i r_i)\right)
\end{equation}
  for the out-of-equilibrium distribution function of the gas.

Following \cite{Stringari_relaxation_time}, we take the moments of
$r_i p_i$ and $p_i^2$ to obtain a closed set of differential equations for the
scaling parameters $\lambda_i$, $\theta_i$:
\begin{align}
  \ddot{\lambda}_i + (2\pi \nu_i)^2 \lambda_i  &-(2\pi \nu_i)^2 \frac{\theta_i}{\lambda_i}  \nonumber \\
  &+ (2\pi \nu_i)^2 \xi_i \left(\frac{\theta_i}{\lambda_i} - \frac{1}{\lambda_i \prod_j \lambda_j}  \right) =0 \,, \label{eq:dyn1}
 \end{align}
 and
 \begin{align}
    \dot{\theta}_i + 2 \frac{\dot{\lambda}_i}{\lambda_i} \theta_i   =-\left(\theta_i -\bar{\theta}\right)/\tau\,,
\end{align}
where $\xi_i = \frac{g/2 (N-1)  \left< n \right> }{g/2 (N-1) \left< n
  \right> +\left<p_i^2 \right>/m}$ accounts for the mean-field interaction with $\left< \cdots \right>$ being phase-space averages with respect to the
equilibrium distribution and $\bar{\theta}=1/3 \sum_i \theta_i$.\\
\textbf{Time-of-flight:} To study the expansion after switching off the trap,
the second term in equation (\ref{eq:dyn1}) is set to 0, and the non-linear coupled
differential equations are solved for the scaling parameters, which when plugged
into the scaling ansatz yield the phase-space distribution after time-of-flight.

For the non-interacting case the equations for the expansion can be solved explicitly. In the collisionless regime, $\theta_i = \lambda_i^{-2}$, and the equation for $\lambda_i$ can be solved to obtain
$\lambda^{(0)}_i(t) = \sqrt{1+(2 \pi \nu_i)^2 t^2}$. Using the non-interacting value
of the initial RMS radii, $\left<x_i^2\right>_0(t=0) = \frac{k_b T}{m (2 \pi \nu_i)^2}$, we
obtain after time-of-flight $\left<x_i^2\right>_0(t) = \frac{k_b T}{m} \frac{1+ (2 \pi \nu_i)^2 t^2}{(2 \pi \nu_i)^2}$, such that the ratio of different directions approaches 1,
and the cloud becomes isotropic.\\
\textbf{Virial expansion for the number fluctuations:} An alternative approach to derive the thermodynamic quantities is provided by
virial expansions of the partition function of the interacting many body
system \cite{Virial1,Virial2,Virial3}.

As we are interested in the number fluctuations, and thus, the isothermal compressibility $\kappa = \frac{1}{n^2}\frac{\partial n}{\partial \mu}$, we start from the expression for the chemical
potential of a homogeneous Fermi gas at low temperature and weak interactions \cite{Virial4}, adapted to the SU$(N)$ case
\begin{equation}
  \begin{split}
  \mu(n,T,a) &= E_F \left[ 1 - \frac{\pi^2}{12} (T/T_F)^2 + \frac{4}{3 \pi} (N-1)k_F a \right. \\
         &\left. + \frac{4(11-2\ln(2))}{15 \pi^2} (k_Fa)^2 (N-1) \right] + C T^2 a^2\,,
  \end{split}
\end{equation}
where $C$ is a constant independent of $n$. We can then evaluate the compressibility from the dependence of $n$ on the Fermi parameters.

For simplicity we only keep terms up to first order, to obtain for the number
fluctuations
\begin{equation}
\Delta \tilde{N}^2/\tilde{N} = k_B T n \kappa =  \frac{3}{2} \frac{T/T_F}{1+\frac{2}{\pi} \,(k_F a) (N-1)} \,,
\end{equation}
which is the equation presented in the main text.

\end{document}